\documentstyle[11pt,AATS2,epsf]{article}	

\begin{document}	

\title{Time Since the Beginning} 

\author{Alan H. Guth}
\affil{Center for Theoretical Physics,
Laboratory for Nuclear Science and Department of Physics,
Massachusetts Institute of Technology, Cambridge,
Massachusetts\ \ 02139\ \ \ U.S.A.}


\begin{abstract}
While there is no consensus about the history of time since the
beginning, in this paper I will discuss some possibilities.  We
have a pretty clear picture of cosmic history from the
electroweak phase transition through the time of recombination, a
period which includes the QCD phase transition and big bang
nucleosynthesis.  This paper includes a quantitative discussion
of the age of the universe, of the radiation-matter transition,
and of hydrogen recombination.  There is much evidence that at
earlier times the universe underwent inflation, but the details
of how and when inflation happened are still far from certain. 
There is even more uncertainty about what happened before
inflation, and how inflation began.  I will describe the
possibility of ``eternal'' inflation, which proposes that our
universe evolved from an infinite tree of inflationary spacetime. 
Most likely, however, inflation can be eternal only into the
future, but still must have a beginning.
\end{abstract}





\def\tot{{\rm tot}}

\section{Introduction}
In this paper I will attempt to discuss the history of time from
the beginning, even though no complete description exists.  In
Sec.~2 I will lay out the basic equations, and in Sec.~3 I will
discuss the time period from about $10^{-12}$ s to 300,000 years. 
In Sec.~4 I will discuss what happened earlier, suggesting that
inflation is the answer.  Sec.~5 will deal with the question of
what happened before inflation, to which I will argue that
the answer is more inflation---i.e., eternal inflation.  In the
final section I will summarize.

\section{Fundamentals of Early Universe Physics}\label{Fundamentals}

The time-evolution of the early universe seems to be
well-described by a remarkably simple theory, known alternatively
as the hot big bang theory or the standard cosmological model. 
The model assumes that the universe is well-approximated as being
homogeneous and isotropic, which implies that the metric can be
written in the Robertson-Walker form, 
\begin{equation}
  {\rm d}s^2 = -{\rm d}t^2 + a^2(t) \left\{ {{\rm d} r^2 \over 1 -
     k r^2} + r^2 \left[ {\rm d}\theta^2 + \sin^2 \theta \, {\rm d}
     \phi^2 \right] \right\} \ ,
  \label{RWmetric}
\end{equation}
where $k$ denotes a constant which indicates whether the universe
is open ($k < 0$), closed ($k > 0$), or flat ($k = 0$), and
throughout this article I will use units for which $\hbar
\equiv c \equiv k_B \equiv 1.$  The Einstein equations imply that
the scale factor $a(t)$ evolves according to
\begin{eqnarray}
  \left({\dot a \over a} \right)^2 &=& {8 \pi \over 3} G
     \rho - {k \over a^2}\label{adot} \label{dyn1}\\
  \ddot a &=& -{4 \pi \over 3} G (\rho + 3 p) a \label{dyn2}\\
  {{\rm d} \over {\rm d}t} \left(a^3 \rho \right) &=& - p \,
     {{\rm d} \over {\rm d}t} \left( a^3 \right) \ , \label{dyn3}
\end{eqnarray}
where $\rho$ is the mass density, $p$ is the pressure, $G$ is
Newton's constant, and the overdot denotes a derivative with
respect to $t$.  These equations are not independent, since any
one of them can be derived from the other two.  Assuming that the
mass density consists of matter ($\rho_m \propto a^{-3}$),
radiation ($\rho_r \propto a^{-4}$), and vacuum mass density
($\rho_{\rm vac} = \hbox{constant}$), then Eq.~(\ref{adot}) can
be integrated to give the relationship between the scale factor
$a$ and the time $t$.  Denoting the present value of the Hubble
parameter $H \equiv \dot a / a$ by $H_0$, and normalizing the
scale factor so that its present value is 1, one finds: 
\begin{equation}
  t = H_0^{-1} \int_0^a {a' \,{\rm d} a' \over \sqrt{\Omega_m a' +
     \Omega_r + \Omega_{\rm vac} a'^4 + \Omega_k a'^2}}
     \ ,\label{age} 
\end{equation}
where $\Omega_X \equiv \rho_{X0}/\rho_c\,$, the subscript $0$
denotes the present time, $\rho_c$ denotes the critical density
$3 H_0^2 / (8 \pi G)$, and $\Omega_k = 1 - \Omega_m - \Omega_r -
\Omega_{\rm vac}$.

There is now much evidence that the universe is flat, coming
predominately from studies of the cosmic microwave background
(CMB)\null.  Wang, Tegmark, \& Zaldarriaga (2001) have carried out a
comprehensive study in which they combined the measurements of
the CMB (most importantly the results from BOOMERaNG (Netterfield
et al.\ 2001), DASI (Halverson et al.\ 2001), Maxima (Lee et al.\
2001), and CBI (Padin et al.\ 2001)) with measurements of large
scale structure (IRAS PSCz survey (Saunders et al.\ 2000;
Hamilton, Tegmark, \& Padmanabhan 2000)) and the Hubble parameter
(Freedman et al.\ 2000) to find that $\Omega_k = 0.0 \pm 0.06$ at
the 95\% confidence level.  Since this result is also in
agreement with the prediction of the simplest inflationary
models, for the remainder of this paper I will consider only
models that are exactly flat ($k=0$). 

\begin{figure}[htb]	
\plotfiddle{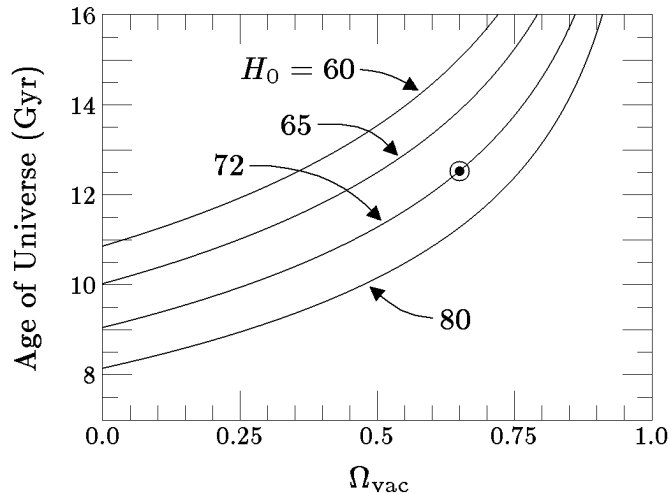}{174pt}{0}{100}{100}{-136}{-26}	
\caption{The calculated age of the universe is shown as a
function of $\Omega_{\rm vac}$, for various values of the Hubble
parameter $H_0$, measured in km-s$^{-1}$-Mpc$^{-1}$.  The models
are flat, and assumed to have negligible radiation density.  The
circled dot shows the currently popular model with $H_0 = 72$,
$\Omega_{\rm vac} = 0.65$, and an age of 12.5 Gyr. \label{figage}}
\end{figure}

To apply Eq.~(\ref{age}) for times near the present, it is
sufficient to neglect $\Omega_r \approx 10^{-4}$, in which case
Eq.~(\ref{age}) can integrated analytically:
\begin{equation}
  t = {2 H_0^{-1} \over 3 \sqrt{\Omega_{\rm vac}}} \, \tanh^{-1}
     \sqrt{ {\Omega_{\rm vac} a^4 \over a^4 \Omega_{\rm vac} + a
     (1 - \Omega_{\rm vac})}} \ . 
  \label{agenorad}
\end{equation}
To determine the present age we set $a=1$ in the above equation,
finding $t_0 = (2 H_0^{-1} / (3 \sqrt{\Omega_{\rm vac}})
\tanh^{-1}\sqrt{\Omega_{\rm vac}}$.  Numerical evaluations of
this formula are shown in Figure \ref{figage}.  The final value
for the Hubble parameter obtained by the Hubble Key Project
(Freedman et al.\ 2000) was $H_0 = 72 \pm 8$
km-s$^{-1}$-Mpc$^{-1}$, so I take $H_0 = 72$ as the central value
in Figure \ref{figage}.  There is more uncertainty in
$\Omega_{\rm vac}$, but I will take $\Omega_{\rm vac} \approx
0.65$ as the central value for the graphs.

While radiation is a negligible contribution to the total mass
density today, the universe is believed to have been
radiation-dominated from just after inflation until some tens of
thousands of years after the big bang.  The energy density of
such thermal radiation is given by
\begin{equation}
  \rho = g {\pi^2 \over 30} T^4 \ ,
  \label{rho}
\end{equation}
where $g$ denotes the number of effectively massless bosonic spin
states, plus 7/8 times the number of effectively massless
fermionic spin states.  The entropy density is given by
\begin{equation}
  s = g {2 \pi^2 \over 45} T^3 \ .
  \label{entropy}
\end{equation}
Except for inflation the entropy of the early universe is
believed to have remained essentially constant, so that the
relationship between time $t$ and temperature $T$ can be found
from $a^3 s = constant$ and the dynamical equations
(\ref{dyn1})--(\ref{dyn3}).  If $g$ can be treated as a constant,
as it can for various time intervals, this relation becomes
\begin{equation}
  T = \left( {45 \over 16 \pi^3 g G} \right)^{1/4} {1 \over
     \sqrt{t}} \ .
  \label{T}
\end{equation}
After the disappearance of the muons at about $10^{-4}$ s, the
contributions to $g$ consist of photons ($g=2$),
electron-positron pairs ($g=7/2$), and three species of neutrinos
($g=21/4$), for a total of $g=10{3 \over 4}$.  During this
interval Eq.~(\ref{T}) reduces to
\begin{equation}
  T = {0.8592 \hbox{ MeV} \over \sqrt{t/(1 \hbox{ sec})}} = {9.971
     \times 10^9 \,\hbox{K} \over \sqrt{t/(1 \hbox{ sec})}} \ . 
  \label{Tprime}
\end{equation}
As an order of magnitude estimate, one can use the above formula
for all times between about $10^{-12}$ s and $10^3$ years.  As a
precise formula, Eq.~(\ref{Tprime}) begins to fail at about $t=1$
s, when the $e^+e^-$ pairs start to disappear from the thermal
equilibrium mix.  By this time the neutrinos have effectively
decoupled, so all the entropy of the $e^+e^-$ pairs (with
$g=7/2$) is given to the photons (with $g=2$) and not the
neutrinos.  As a result, the entropy density of photons is
increased by a factor of $({7 \over 2}+2)/2 = 11/4$,
so the temperature of the photons relative to the neutrinos is
increased by a factor of $(11/4)^{1/3}$.  This ratio is believed
to persist to the present.

For the period after the disappearance of the $e^+e^-$ pairs, one
conventionally uses $T$ to denote the temperature of the photons,
while the neutrinos have a temperature $T_\nu=(4/11)^{1/3}\, T$. 
The COBE FIRAS measurements (Mather et al.\ 1999) determined that
$T_0=2.725 \pm 0.002\,$K\null, which implies a present mass
density in photons and neutrinos of $7.804 \times 10^{-34} \hbox{
g/cm}^3$.  Using $H_0 = 72 \hbox{ km-s$^{-1}$-Mpc$^{-1}$}$, one
finds $\Omega_r = 8.013 \times 10^{-5}$. 

\section{Cosmic Events from $10^{-12}$ Second to 300,000
Years}\label{classical}

The first key event of this period is the electroweak phase
transition, at which the SU(2)$\times$U(1) symmetry of the
Glashow-Weinberg-Salam electroweak theory is broken to the
familiar U(1) symmetry of quantum electrodynamics.  At this phase
transition a Higgs field is believed to acquire a nonzero
expectation value.  The interaction of the Higgs field with other
fields is then responsible for masses of the corresponding
particles, which include the $W$, the $Z$, the leptons ($e$,
$\mu$, and $\tau$), and the quarks.  It is worth noting, however,
that the masses acquired by the $u$ and $d$ quarks through the
electroweak symmetry-breaking are called the ``current-quark''
masses, and have values under 10 MeV\null.  They have very little
influence on the masses of protons and neutrons, which are
associated with the ``constituent'' quark masses that arise from
the strong interactions of the quarks.

The details of the electroweak phase transition remain unknown,
since the Higgs particle and its detailed properties remain out
of reach.  The lack of attractive alternatives has convinced most
particle physicists that the Higgs particle almost certainly
exists, but it remains possible that nature is more complicated
than the simple models with a single Higgs field.  The energy
scale of electroweak symmetry breaking is certainly, however, on
the order of 1 TeV, so the time of the electroweak phase
transition can be estimated from Eq.~(\ref{Tprime}) at about
$10^{-12}$~s.

The next important event was the quantum chromodynamics (QCD)
phase transition, which has an energy scale of approximately 1
GeV, and therefore took place at about $10^{-6}$ s.  At this
phase transition the quark-gluon plasma, with its essentially
free quarks, disappeared in favor of a phase in which the quarks
are permanently bound inside mesons and baryons.  At about the
same time the overwhelming majority of quarks and antiquarks
annihilated in pairs.  A tiny excess of quarks over antiquarks, of
about one part in $10^9$, resulted in the survival of a tiny
fraction of the hadronic matter, and this tiny excess is
responsible for the existence of the protons and neutrons that
populate the current universe.  We believe that the excess was
generated by a process, known as baryogenesis, which may have
occurred anytime from the grand unified theory era through the
electroweak phase transition. 

At $t \approx 1$ s, when the temperature fell to about 1 MeV, the
processes that led to big bang nucleosynthesis began.  The first
step was the decoupling of the neutrinos, which cut off the
reactions that had until this time maintained a thermal
equilibrium balance between protons and neutrons.  Since the
neutron mass exceeds the proton mass by 1.29 MeV, the number of
neutrons was suppressed relative to the number of protons, but not
by a large amount.  

It is often pointed out that it appears to be an important
coincidence that the temperature at which the neutrinos decouple,
determined by the strength of the weak interactions and various
cosmological parameters, is very nearly equal to the
neutron-proton mass difference, which is presumably the result of
an interplay between the strong and electromagnetic interactions. 
If the neutrinos remained coupled for much longer, the thermal
equilibrium between protons and neutrons would be maintained down
to lower temperatures, resulting in the almost complete
disappearance of neutrons from the universe.  If the neutrinos
decoupled earlier, then the universe would be left with a nearly
50\%/50\% mix of protons and neutrons, which would result in an
almost total conversion to He$^4$ in big bang nucleosynthesis. 

After the neutrinos decoupled at $t \approx 1$ s, the only
relevant reaction that could interchange protons and neutrons was
the free decay of the neutron, with a mean life of about 15
minutes.  After about 3 to 4 minutes, however, the temperature
fell to $T \approx 0.1$ MeV, which is cool enough for the
deuteron to become stable.  At this point nuclear reactions
proceeded quickly, converting almost all the neutrons that
remained into He$^4$, which today has an abundance of 23\%--25\%
by mass (see, for example, Burles, Nollett, \& Turner 2001b).  In
addition, detectable amounts of deuterium, He$^3$, and Li$^7$
were produced.  Note that 0.1 MeV is far below the deuteron
binding energy of 2.2 MeV, but that such low temperatures are
needed for stability because of the huge ratio of photons to
baryons, about $10^9:1$.  Thus each deuteron that formed must
have survived a huge number of photon collisions before it had
the chance to proceed with further nuclear reactions.

\begin{figure}[htb]	
\plotfiddle{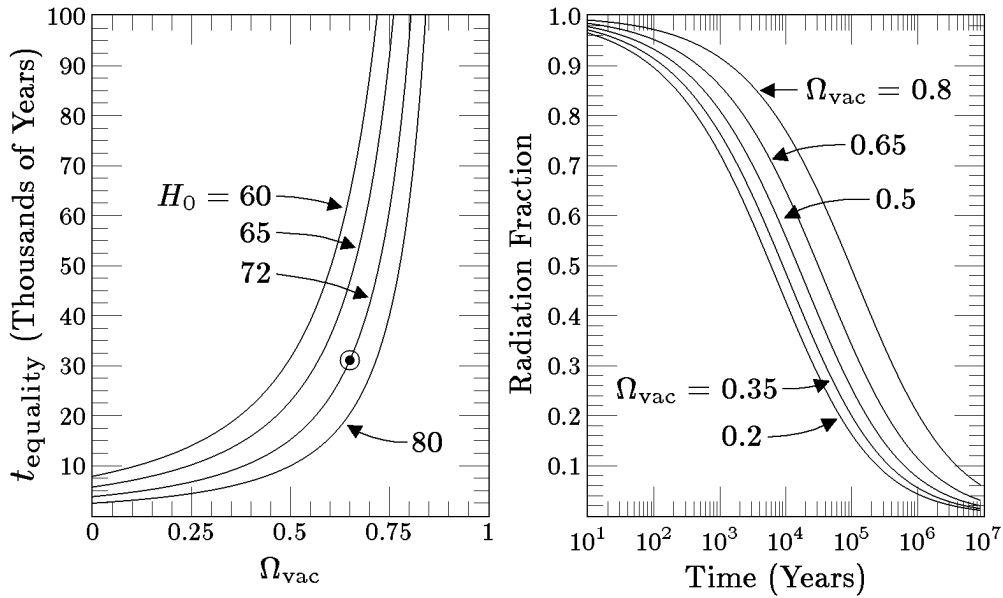}{216pt}{0}{100}{100}{-189}{-16}	
\caption{The left graph shows the time of matter-radiation
equality as a function of the present value of $\Omega_{\rm
vac}$, for various values of $H_0$ (in km-s$^{-1}$-Mpc$^{-1}$). 
The circled dot shows the currently popular model with $H_0 = 72$
and $\Omega_{\rm vac} = 0.65$, with $t_{\rm equality} = 31,070$
yr.  The graph on the right shows the fraction of the total mass
density of the universe in radiation as a function of time, for
various values of $\Omega_{\rm vac}$.  Both graphs represent the
same flat cosmological models, which are assumed to have three
species of massless neutrinos and a present radiation temperature
of 2.725$\,$K\null.  The ``dark energy'' component is taken to be
a cosmological constant, with a fixed vacuum mass density.
\label{figtequality}}
\end{figure}

At $t \approx 30,000$ years, the mass density of the universe
gradually changed from radiation-dominated to matter-dominated,
where ``matter'' refers to both dark matter and baryonic matter. 
This change is described by the equations presented in Sec.~2,
and is actually a very gradual transition.  The results of a
numerical integration of these equations is shown in Figure
\ref{figtequality}. 

Finally, the last important event of this period is known as
hydrogen ``recombination,'' although ``combination'' would be a
more accurate term.  In the context of the standard cosmological
model, the electrons and protons had never been combined at any
point in the past.  Recombination is often said to take place at
a temperature of 4000$\,$K and at a time of 300,000 years.  These
numbers are in fact reasonable estimates, but the actual process
of recombination, like that of matter-domination, is gradual. 
Note that $4000\,\hbox{K} \approx 0.34$ eV, so like the deuteron
during nucleosynthesis, atomic hydrogen in the early universe did
not become stable until $k_B T$ was far below its binding energy.

If one assumes thermal equilibrium, then the fraction $x$ of
protons or electrons that remain ionized is given by the Saha
equation, 
\begin{equation}
  {x^2 \over 1-x} = {(2 \pi m_e k_B T)^{3/2} \over (2 \pi
     \hbar)^3 \, n}\, e^{-B/k_B T} \ .
\end{equation}
Here $m_e$ is the electron mass, $k_B$ is the Boltzmann constant,
and $B$ is the binding energy of hydrogen, 13.60 eV\null.  (For a
pedagogical treatment of the Saha equation, see Peebles 1993,
pp.~165--167.)  The Saha equation provides a reasonable
approximation for the onset of recombination, but the process
soon departs significantly from thermal equilibrium, as was shown
by Peebles (1968).  The reasons for the departure from thermal
equilibrium are a bit subtle, since the reaction rates for
ionization and recombination are much faster than the expansion
rate of the universe.  The problem, however, is that almost every
decay to the ground state of hydrogen emits a Lyman alpha photon
which then has a high probability of ionizing another hydrogen
atom.  Thus, the sum of the number of ground state hydrogen atoms
plus the number of Lyman alpha photons changes slowly, and lags
behind thermal equilibrium as the universe cools.  The dominant
mechanisms for changing this sum are the rare two-photon decay of
the 2s level of hydrogen to the ground state, and the gradual
redshifting of the Lyman alpha photons out of the relevant range
of frequencies. 

\begin{figure}[htb]	
\plotfiddle{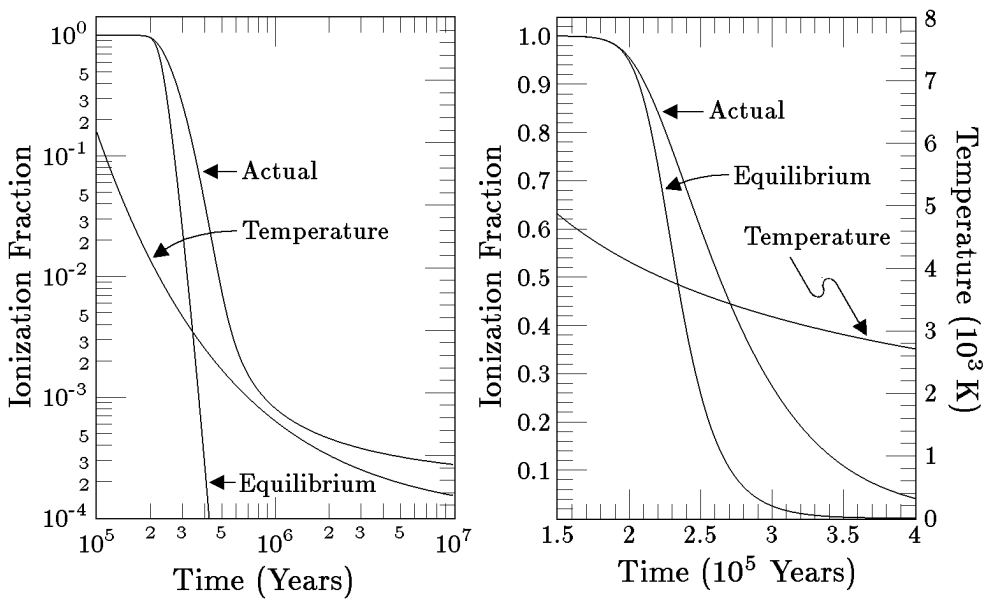}{216pt}{0}{100}{100}{-189}{-16}	
\caption{Both graphs show the process of recombination as a
function of time, for a model described in the text.  The
ionization fraction is the fraction of all protons or electrons
that are ionized at any given time.  The left graph is
logarithmic, and the right graph is linear.  The line labeled
``Equilibrium'' shows the result of solving the Saha equation,
while the line labeled ``Actual'' shows the result of integrating
the rate equations, showing the nonequilibrium effects.  For
reference, the temperature is also shown, keyed to the scale on
the right which applies to both graphs.
\label{figrecomb}}
\end{figure}

Numerical results for recombination are shown in Figure
\ref{figrecomb}, using the currently indicated values of the
parameters.  In particular, the calculations use a flat model
with $T_0 = 2.725\,$K, $\Omega_{\rm vac} = 0.65$, and $\Omega_B
h^2 = 0.020$, following Burles, Nollett, \& Turner (2000a \& b),
who found $\Omega_B h^2 = 0.020 \pm 0.002$ (95\% confidence
level).  (Note that $h$ is defined by $H_0 = 100h$
km-s$^{-1}$-Mpc$^{-1}$.)  The results were obtained by numerical
integrations carried out by me, using the equations of Peebles
(1968) and Peebles (1993), pp.~165--173.

\section{Before $10^{-12}$ Second: Inflation}

At times before $10^{-12}$ second, some of us believe that the
universe almost certainly underwent a period of inflation.  The
reason is that cosmic inflation can explain a number of features
of our universe that would otherwise be unexplained.  In
particular, inflation can explain:

\begin{enumerate}
\item{\it How the universe acquired $> 10^{90}$ particles}

Starting from the general and then moving toward the specific,
one salient feature of the universe is its enormous size. 
The visible part of the universe contains about $10^{90}$
particles.  It is easy to take this for granted, and many
cosmologists are not bothered by the fact that the ``standard''
FRW cosmology, without inflation, simply postulates that about
$10^{90}$ or more particles were here from the start.  However,
in the present context many of us hope that even the creation of
the universe can be described in scientific terms, and thus the
number of particles would have to be the result of some
calculation.  The easiest way by far to get a huge number, with
presumably only modest numbers as input, is for the calculation
to involve an exponential.  The exponential expansion of
inflation reduces the problem of explaining $10^{90}$ particles
to the problem of explaining 60 to 70 e-foldings of inflation. 
In fact it is easy to construct underlying particle theories
that will give far more than 70 e-foldings of inflation, so
inflationary cosmology suggests that the observed universe is
only an infinitesimal fraction of the entire universe.

\item{\it Why the universe is uniformly expanding}

The Hubble expansion is also easy to take for granted, and in the
standard FRW cosmology the Hubble expansion is accepted as a
postulate about the initial conditions.  But inflation offers the
possibility of actually explaining how the Hubble expansion
began.  The repulsive gravity associated with the inflaton
field---the scalar field that drives the inflation---is exactly
the kind of force needed to propel the universe into a pattern of
motion in which each pair of particles is moving apart with a
velocity proportional to their separation.

\item{\it How the CMB can be uniform to 1 part in $10^{5}$}

The degree of uniformity in the universe is startling.  The
intensity of the cosmic background radiation is the same in all
directions, after it is corrected for the motion of the Earth, to
the incredible precision of one part in 100,000.  

The cosmic background radiation was released at the time of
recombination, about 300,000 years after the big bang, when the
universe cooled enough so that the opaque plasma neutralized into
a transparent gas.  The cosmic background radiation photons have
mostly been traveling on straight lines since then, so they
provide an image of what the universe looked like at 300,000
years after the big bang.  The observed uniformity of the
radiation therefore implies that the observed universe had become
uniform in temperature by that time.  In standard FRW cosmology,
a simple calculation shows that the uniformity could be
established so quickly only if signals could propagate at 100
times the speed of light, a proposition clearly in contradiction
with the known laws of physics.  In inflationary cosmology,
however, the uniformity is easily explained.  The uniformity is
created initially on microscopic scales, by normal
thermal-equilibrium processes, and then inflation takes over and
stretches the regions of uniformity to become large enough to
encompass the observed universe.

\item{\it Why the early universe was so close to critical
density}

I find this issue particularly impressive, because of the
extraordinary numbers that it involves. This ``flatness problem''
concerns the value of the ratio
\begin{equation}
  \Omega_\tot \equiv {\rho_\tot \over \rho_c} \ ,
  \label{omegatot}
\end{equation}
where $\rho_\tot$ is the average total mass density of the
universe and $\rho_c = 3 H^2 / (8 \pi G)$ is the critical density,
the density that would make the universe spatially flat.  (In the
definition of ``total mass density,'' I am including the vacuum
energy $\rho_{\rm vac} = \Lambda/ 8 \pi G$ associated with the
cosmological constant $\Lambda$, if it is nonzero.)

There is now strong evidence that $\Omega$ is very near to 1, but
the flatness problem is much older and does not require us to
believe the most recent results.  We have believed for a long
time that 
\begin{equation}
  0.1 \la \Omega_0 \la 2 \ ,
  \label{omegabounds}
\end{equation}
and this is all that is needed to motivate the flatness problem. 
Despite the breadth of this range, the value of $\Omega$ at early
times is highly constrained, since $\Omega=1$ is an unstable
equilibrium point of the standard model evolution.  Thus, if
$\Omega$ was ever {\it exactly} equal to one, it would remain
exactly one forever.  However, if $\Omega$ differed slightly from
one in the early universe, that difference---whether positive or
negative---would be amplified with time.  In particular, it can
be shown that $\Omega - 1$ grows as
\begin{equation}
  \Omega - 1 \propto \cases{t &(during the radiation-dominated era)\cr
    t^{2/3} &(during the matter-dominated era)\ .\cr}
  \label{omegagrowth}
\end{equation}
It was shown by Dicke and Peebles (1979), for example, that as
the processes of big bang nucleosynthesis were just beginning at
$t=1$ sec, $\Omega$ must have equaled one to an accuracy of one
part in $10^{15}$.  Classical cosmology provides no explanation
for this fact---it is simply assumed as part of the initial
conditions.  In the context of modern particle theory, where we
try to push things all the way back to the Planck time,
$10^{-43}$ sec, the problem becomes even more extreme.  If one
specifies the value of $\Omega$ at the Planck time, it has to
equal one to 58 decimal places in order to be anywhere in the
allowed range today. 

While this extraordinary flatness of the early universe has no
explanation in classical FRW cosmology, it is a natural
prediction for inflationary cosmology.  During the inflationary
period, instead of $\Omega$ being driven away from one as
described by Eq.~(\ref{omegagrowth}), $\Omega$ is driven towards
one with exponential swiftness:
\begin{equation}
  \Omega - 1 \propto e^{-2 H_{\rm inf} t} \ ,
  \label{omegainflate}
\end{equation}
where $H_{\rm inf}$ is the Hubble parameter during inflation. 
Thus, as long as there is a long enough period of inflation,
$\Omega$ can start at almost any value, and it will be driven to
one by the exponential expansion. 

\item{\it Why the inhomogeneities have a nearly flat
(Harrison-Zeldovich) spectrum}

The process of inflation smooths the universe essentially
completely, but density fluctuations are generated as inflation
ends by the quantum fluctuations of the inflaton field, the
scalar field that drives the inflationary expansion.  Generically
these are adiabatic Gaussian fluctuations with a nearly
scale-invariant spectrum (Starobinsky 1982; Guth \& Pi 1982;
Hawking 1982; Bardeen, Steinhardt, \& Turner 1983; Mukhanov,
Feldman, \& Brandenberger 1992).  New data is arriving quickly,
but so far the observations are in excellent agreement with the
predictions of the simplest inflationary models.  For a review,
see for example Bond and Jaffe (1999), who find that the combined
data give a slope of the primordial power spectrum within 5\% of
the preferred scale-invariant value.  See also Wang, Tegmark, \&
Zaldarriaga (2001), which includes a review of the most current
data. 

\end{enumerate}

Since the theme here is time and time scales, it is natural to
ask {\it when} inflation occurred.  The answer is that we do not
really know.  Originally inflation was proposed to take place at
the scale of grand unified theories, at a characteristic energy
scale of $10^{16}$ GeV (Guth 1981; Linde 1982; Albrecht \&
Steinhardt 1982).  Applying Eq.~(\ref{age}) with $g \approx 200$,
typical of grand unified theories, one finds a starting time for
inflation of about $10^{-39}$ s.  This is extraordinarily early,
but it is still late compared to the Planck time, $\sqrt{G \hbar
/ c^5} \approx 5 \times 10^{-44}$ s, the time scale at which
quantum gravity is believed to become important.  Thus, it is
plausible that the field theoretic formalism that is used to
describe inflation is valid at the appropriate energy scale. 

It is possible that inflation did occur at the grand unified
theory scale, but it might very well have occurred later.  The
only known restriction on the lateness of inflation is the
requirement that baryogenesis occur after inflation, since any
net density of baryon number generated before inflation would be
diluted to a negligible level.  It is now believed that
baryogenesis might happen as late as the electroweak scale, 
operating through the mechanism of electroweak current
conservation anomalies (Kuzmin, Rubakov, \& Shaposhnikov 1985). 

Observationally it is difficult to determine the energy scale and
hence the time scale of inflation, since the consequences are
very insensitive.  The only known way to determine the energy
scale of inflation is to directly or indirectly measure the
gravitational wave background, which is more intense if the
energy scale of inflation was high.  In fact the energy scale
could not have been significantly higher than the grand unified
scale, or else the gravity waves would be so strong that they
should have already been detected.

\section{Before Inflation: (Eternal) Inflation}

The question of what happened before inflation is an open one,
and different cosmologists would venture different ideas.  In my
opinion, the most plausible answer to what happened before
inflation is --- more inflation.

Specifically, it appears that essentially all working models of
inflation are eternal, in the sense that once inflation starts,
it never stops.  Instead inflation goes on forever, with pieces
of the inflating region breaking off and producing a never-ending
stream of ``pocket universes'' (Vilenkin 1983; Steinhardt 1983;
Linde 1986a\&b; Goncharov, Linde, \& Mukhanov 1987).

The mechanism that leads to eternal inflation is rather
straightforward to understand.  Normally one expects inflation to
end because the ``false vacuum''---the state of the inflaton
field that is responsible for the repulsive gravity driving the
inflation---is unstable, so it decays like a radioactive
substance.  As with familiar radioactive materials, the decay of
the false vacuum is generally exponential: during any period of
one half-life, on average half of it will decay.  This case is
nonetheless very different from familiar radioactive decays,
however, because the false vacuum is also expanding
exponentially.  Furthermore, it turns out that the expansion is
generally much faster than the decay.  Thus, if one waits for one
half-life of the decay, half of the false vacuum region would on
average convert to ordinary matter.  But meanwhile the part that
remains would have undergone many doublings, so it would be much
larger than the region was at the start.  Even though the false
vacuum is decaying, the volume of the false vacuum would actually
grow with time.  The volume of the false vacuum would continue to
grow, without limit and without end.  Meanwhile pieces of the
false vacuum region decay, producing an infinite number of what I
call pocket universes.

\begin{figure}[htb]	
\plotfiddle{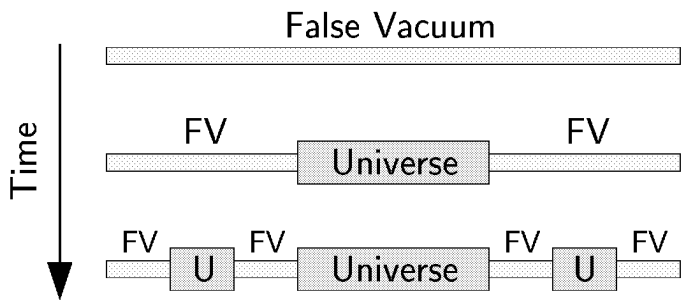}{96pt}{0}{100}{100}{-144}{0}	
\caption{An illustration of eternal inflation, as described in
the text.
\label{figeternal}}
\end{figure}

In Figure \ref{figeternal} I show a schematic illustration of how
this works.  The top row shows a region of false vacuum, shown
very schematically as a horizontal bar.  After a certain length
of time, a little less than a half-life, the situation looks like
the second bar, in which about a third of the region has decayed. 
The energy released by that decay produces a pocket universe,
which will inflate to become much larger than the presently
observed universe.

On the second bar, in addition to the pocket universe, there are
two regions of false vacuum.  On the diagram I have not tried to
show the expansion, so the diagram can fit on the page.  So, you
are expected to remember that each bar is actually bigger than
the previous bar, but drawn on a smaller scale so that it looks
the same size.  To discuss a definite example, let us assume that
each bar represents three times the volume of the previous bar. 
In that case, each region of false vacuum on the second bar is
just as big as the entire bar on the top line.

The process can then repeat.  If we wait the same length of time
again, the situation will be as illustrated on the third bar of
the diagram, which represents a region 3 times larger than the
second bar, and 9 times larger than the top bar. For each region
of false vacuum on the second bar, about a third of the region
decays and becomes a pocket universe, leaving regions of false
vacuum in between.  Each region of false vacuum shown on the
diagram is as large as the original region in the top bar.  The
process goes on literally forever, producing pocket universes and
regions of false vacuum between them, ad infinitum.  The universe
on the very large scale acquires a fractal structure. 

     The illustration of Figure \ref{figeternal} is of course
oversimplified in a number of ways: it is one-dimensional instead
of three-dimensional, and the decays are shown as if they were
very systematic, while in fact they are random.  But the
qualitative nature of the evolution is nonetheless accurate:
eternal inflation really leads to a fractal structure of the
universe, and once inflation begins, an infinite number of pocket
universes are produced.

     Since inflation is eternal into the future, it is natural to
ask if it might also be eternal into the past.  The explicit
models that have been constructed are eternal only into the
future and not into the past, but that does not show whether or
not is possible for inflation to be eternal into the past.  Borde
\& Vilenkin (1994) presented a proof that an eternally inflating
spacetimes must start from an initial singularity, and hence must
have a beginning, but later they pointed out (1997) that their
proof assumed a condition that is true classically but is
violated by quantum field theories.  Today the issue is
undecided.  My own suspicion is that eternally inflating
spacetimes must have initial singularities, because it seems 
significant that no one has been able to construct a model which
does not.

\section{Summary}
For the period between about $10^{-12}$ s and 300,000 years, we
have a rather detailed description of cosmology that I believe
has a good chance of being correct.  I believe that inflation
played a very significant role at earlier times, but the details
are unclear.  As one might expect, our view of the earliest
moments of the universe is still clouded with uncertainties.


\acknowledgments
This work is supported in part by funds provided by the U.S.
Department of Energy (D.O.E.) under cooperative research
agreement \#DF-FC02-94ER40818.


\begin{references}

\reference
Albrecht, A. \& Steinhardt, P. J. 1982, \prl, 48, 1220

\reference
Bardeen, J. M., Steinhardt, P. J., \& Turner, M. S. 1983, \prd,
28, 679

\reference
Bond, J. R., \& Jaffe, A. H. 1999, Phil.Trans.Roy.Soc.Lond.A,
357, 57, astro-ph/9809043

\reference
Borde, A. \& Vilenkin, A. 1994, \prl, 72, 3305, gr-qc/9312022

\reference
Borde, A. \& Vilenkin, A. 1997, \prd, 56, 717, gr-qc/9702019

\reference
Burles, S., Nollett, K. M., \& Turner, M. S. 2001a, \prd, 63,
063512, astro-ph/0008495

\reference
Burles, S., Nollett, K. M., \& Turner, M. S. 2001b, astro-ph/0010171

\reference 
Dicke, R. H., \& Peebles, P.J.E. 1979, in General Relativity: An
Einstein Centenary Survey, ed.\ S.~W.~Hawking \& W.~Israel
(Cambridge: Cambridge University Press), 504

\reference
Freedman, W. L. et al. 2000, astro-ph/0012376, to be published in
\apj

\reference
Goncharov, A. S., Linde, A. D., \& Mukhanov, V. F. 1987,
Int.J.Mod.Phys, A2, 561

\reference
Guth, A. H. 1981, \prd, 23, 347

\reference
Guth, A. H., \& Pi, S.-Y. 1982, \prl, 49, 1110

\reference
Halverson, N. W. et al. 2001, astro-ph/0104489

\reference
Hamilton, A.J.S., Tegmark, M., \& Padmanabhan, N. 2000, \mnras,
317, L23, astro-ph/0004334

\reference
Hawking, S. W. 1982, Phys.Lett, 115B, 295

\reference
Lee, A. T. et al. 2001, astro-ph/0104459

\reference
Linde, A. D. 1982, Phys.Lett, 108B, 389

\reference
Linde, A. D. 1986a, Mod.Phys.Lett., A1, 81

\reference
Linde, A. D. 1986b, Phys.Lett, 175B, 395

\reference
For a modern review, see
Mukhanov, V. F., Feldman, H. A., \& Brandenberger, R. H. 1992, 
Phys.Rep, 215, 203

\reference
Netterfield, C. B. et al. 2001, astro-ph/0104460

\reference
Padin, S. et al. 2001, ApJL, 549, L1, astro-ph/0012211

\reference
Peebles, P.J.E. 1968, \apj, 153, 1

\reference
Peebles, P.J.E. 1993, Principles of Physical Cosmology
(Princeton: Princeton University Press)

\reference
Kuzmin, V. A., Rubakov, V. A., \& Shaposhnikov, M. E. 1985,
Phys.Lett, 155B, 36

\reference
Saunders, W. et al. 2000, \mnras, 317, 55, astro-ph/0001117

\reference
Starobinsky, A.~A. 1982, Phys.Lett, 117B, 175

\reference
Steinhardt, P. J. 1983, in The Very Early Universe, Proceedings
of the Nuffield Workshop, Cambridge, 21 June -- 9 July, 1982, ed.\
G.W.~Gibbons, S.W.~Hawking, \& S.T.C.~Siklos (Cambridge:
Cambridge University Press), 251

\reference
Vilenkin, A. 1983, \prd, 27, 2848

\reference
Wang, X., Tegmark, M., \& Zaldarriaga, M. (2001), astro-ph/0105091

\end{references}
\end{document}